\documentclass[lettersize,journal]{IEEEtran}
\usepackage{amsmath,amsfonts}
\usepackage{algorithmic}
\usepackage{algorithm}
\usepackage{array}
\usepackage[caption=false,font=normalsize,labelfont=sf,textfont=sf]{subfig}
\usepackage{textcomp}
\usepackage{stfloats}
\usepackage{url}
\usepackage{verbatim}
\usepackage{graphicx}
\usepackage{cite}
\usepackage{makecell}  
\begin{document}

\title{Topology-Aware Cooperative Beam-Hopping Scheduling for Efficient Resource Allocation in LEO Satellite Systems}

\author{Shuang Zheng, Xing Zhang,~\IEEEmembership{Senior Member,~IEEE,} Quan Z. Sheng, and Wenbo Wang,~\IEEEmembership{Senior Member,~IEEE}

\thanks{(Corresponding authors: Xing Zhang.)}
\thanks{Shuang Zheng is with School of Electronic and Information Engineering, Tiangong University, Tianjin 300387, China (e-mail:
zshuang@tiangong.edu.cn.}
\thanks{ Xing Zhang, and Wenbo Wang
are with the School of Information and Communication Engineering, Beijing
University of Posts and Telecommunications, Beijing 100876, China. (e-mail:\{hszhang; wbwang\}@bupt.edu.cn).}
\thanks{Quan Z. Sheng is with Beijing Normal-Hong Kong Baptist University
 Zhuhai, China. (e-mail: michaelsheng@bnbu.edu.cn).}
}

\markboth{IEEE Wireless Communications Letters,~Vol.~xx, No.~x, April~2026}%
{Shell \MakeLowercase{\textit{et al.}}: A Sample Article Using IEEEtran.cls for IEEE Journals}

\IEEEpubid{0000--0000/00\$00.00~\copyright~2026 IEEE}

\maketitle

\begin{abstract}
The dynamic topology and heterogeneous traffic demands of multi-satellite systems present substantial challenges for beam-hopping (BH) resource management. This letter develops a topology-aware cooperative BH framework for multi-satellite systems under partial observability. A lightweight load-balancing scheme first determines the serving relationships, which are then exploited to construct an association-induced satellite-user graph. A two-phase graph neural network (GNN) extracts episode-level topology-aware structural identities for parameter-sharing agents, while a recurrent multi-agent proximal policy optimization (MAPPO) policy handles slot-level dynamics for joint beam scheduling and onboard power control. Simulation results demonstrate notable gains in energy efficiency, throughput, and fairness over the baselines.
\end{abstract}


\section{Introduction}
\IEEEPARstart{W}{ith} the rapid deployment of large low-Earth-orbit (LEO) constellations \cite{1}, satellite communication systems are advancing toward high-capacity and highly flexible architectures \cite{2}. The strong satellite mobility and the spatio-temporal heterogeneity of traffic impose stringent requirements on dynamic resource scheduling. Beam-hopping (BH) offers flexible beam adaptation to meet diverse traffic demands. However, conventional single-satellite BH schemes cannot fully exploit constellation-wide resources, highlighting the need for inter-satellite cooperative scheduling.

In multi-satellite cooperative scenarios, BH faces several key challenges. First, the rapidly evolving constellation topology leads to highly dynamic and tightly coupled coverage overlaps and interference patterns. Second, during operation, each satellite can only access local observations, which severely limits the scalability and real-time applicability of centralized scheduling methods.

Recently, multi-satellite beam scheduling has received growing research interest. In \cite{3}, a single-satellite BH framework was extended to support multi-satellite cooperation for interference avoidance, achieving low computational complexity but still limited by restricted inter-satellite information exchange. With the development of AI technologies, several works have applied single-agent deep reinforcement learning (DRL) for resource allocation \cite{5}, \cite{6}, where a gateway performed centralized decision-making based on global states. However, the latency and signaling overhead of satellite–ground interactions inevitably reduced scheduling flexibility. To enhance distributed cooperation, the studies in \cite{4} and \cite{7} adopted multi-agent DRL under a centralized training and decentralized execution (CTDE) paradigm to enable cooperative BH among multiple satellites. Nonetheless, their action space modeling remained restrictive, limiting scalability to large-scale scheduling scenarios. Moreover, Ren et al.  \cite{8} proposed a multi-agent cooperative BH scheme for large LEO constellations, but the approach could not explicitly incorporate network topology information, which constrains the representational capability of the system states.

To address these challenges, we propose a topology-aware cooperative beam scheduling framework for multi-satellite systems. A low-complexity association algorithm first balances heterogeneous traffic among visible satellites, and the resulting serving relationships are used to construct a satellite–user graph. A two-phase GNN then transforms the association-induced topology into topology-aware satellite representations, which provide structural identities for agents using a parameter-sharing policy. Based on these representations, a recurrent MAPPO policy performs decentralized joint beam scheduling and power control under the CTDE framework.

\section{System Model and Problem Formulation}

\subsection{System Model}

\begin{figure}[!t]
\centering
\includegraphics[width=0.7\columnwidth]{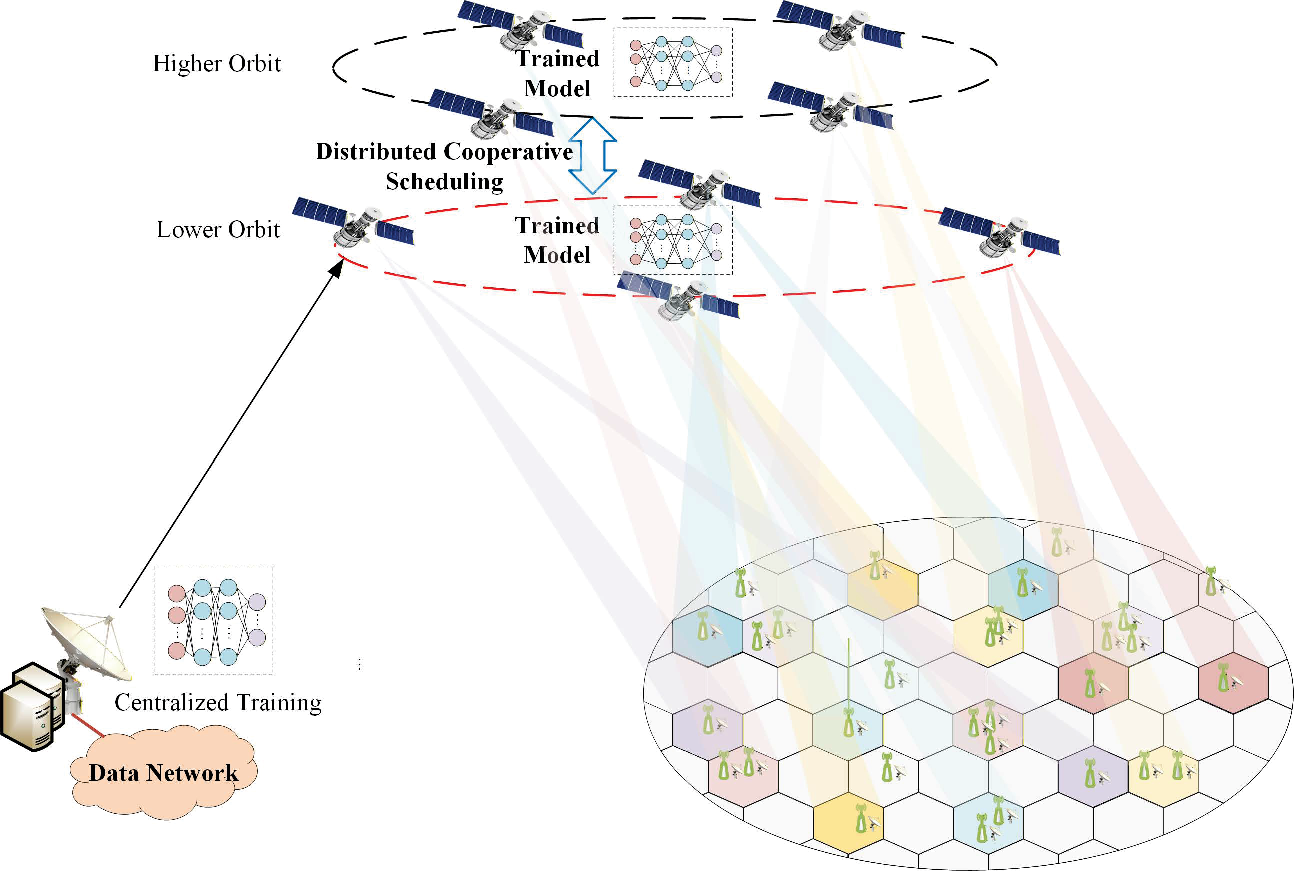}
\vspace{-2mm}
\caption{The architecture of coordinated multi-satellite beam-hopping.}
\label{fig_1}
\vspace{-5mm}
\end{figure}
\IEEEpubidadjcol

As shown in Fig.~\ref{fig_1}, we focus on a multi-LEO-satellite cooperative BH forward-link scenario over consecutive $T_s$ time slots. Within the considered region, $N_u$ users are distributed across $N_m$ beam positions, which are covered by $N_s$ LEO satellites. Let $\mathcal{M}_s$ and $\mathcal{U}_s$ denote the sets of beam positions activated by and users associated with the $s$-th satellite, respectively. At time slot $t$, the number of active beams per satellite is limited to $K\ (K \le  |{{\cal M}_s}|)$.

Considering the heterogeneity of traffic demand distributions, the traffic of the $u$-th user transmitted by the $s$-th satellite follows a Poisson arrival process with arrival rate $\lambda_{u,s}$. The onboard buffer can store only the traffic packets arrived in the last $T_{ttl}$ time slots, where $T_{ttl}$ denotes the time to live (TTL). 

Assume that each beam is associated with at most one beam position. Users in hotspot areas are equipped with two receiving antennas, whereas users outside hotspot regions use a single antenna. A widely adopted channel model is considered 
and $P_{u}^{s,k,t}$ represents the transmit power allocated to the $u$-th user on the $k$-th beam of the $s$-th satellite. It is assumed that the onboard transmit power is uniformly allocated among all active beams and is further evenly distributed among the subchannels within each beam. To enhance the spectral efficiency of the LEO satellite BH system, full frequency reuse among beams is adopted.
Accordingly, the signal-to-interference-plus-noise ratio (SINR) experienced by the $u$-th user from the $k$-th beam of the $s$-th satellite at 
$t$ is given by:
\begin{equation}
\label{eq1}
\begin{aligned}
&\mathrm{SINR}_{u}^{s,k,t}= \\
&
\frac{{P_u^{s,k,t}{{\left| {h_u^{s,k,t}} \right|}^2}}}{{\sigma {{_u^{s,k,t}}^2} + \sum\limits_{k' = 1\atop
k' \ne k}^K {P_u^{s,k',t}} {{\left| {h_u^{s,k',t}} \right|}^2} + \sum\limits_{s' = 1\atop
s' \ne s}^{{N_s}} {\sum\limits_{k = 1}^K {P_u^{s',k,t}} } {{\left| {h_u^{s',k,t}} \right|}^2}}}.
\end{aligned}
\end{equation}

In Eq. \eqref{eq1}, $\sigma {{_u^{s,k,t}}^2}$ denotes the additive white Gaussian noise power, and the co-channel interference comprises both intra-satellite and inter-satellite interference. In multi-satellite multi-beam systems, beam management is therefore a key design challenge. Accordingly, the achievable data rate of user $u$ from 
the $k$-th beam of satellite $s$ at time slot $t$ is given by:
\begin{equation}
\label{eq2}
R_u^{s,k,t} = \mathbb{I}(x_u^{s,k,t}= 1)B_u^{s,k,t}{\log _2}(1 + SINR_u^{s,k,t}),
\end{equation}
where $\mathbb{I}(\cdot)$ denotes the indicator function and $B_u^{s,k,t}$ is the allocated bandwidth, and the corresponding throughput is $Th_u^{s,k,t} = \min \{ R_u^{s,k,t}{T_{slot}},q_{s,u}^t\}$.

\vspace{-5mm}
\subsection{Problem Formulation} 
Based on the above 
model, the beam scheduling problem in the multi-satellite cooperative BH LEO satellite system is decomposed into two phases, while the proposed solution framework is developed in Section III. In the first phase, a user–satellite association subproblem is considered to determine a serving satellite for each user, which serves as a pre-processing step prior to beam scheduling.
In the second phase, given the established user association relationships, a multi-satellite multi-beam cooperative beam scheduling subproblem is investigated, in which coordinated decisions among multiple LEO satellites are exploited to enhance the overall system performance.

\subsubsection{User-satellite association subproblem} 
In multi-satellite cooperative BH LEO systems, user–satellite association critically affects subsequent beam scheduling. Excessive user concentration on a subset of satellites results in load imbalance, which degrades the performance of cooperative scheduling. Accordingly, a preliminary load balancing among satellites is required prior to beam scheduling, and the corresponding subproblem can be formulated as minimizing the traffic load difference among satellites. Specifically:
\begin{equation}
\label{eq3}
\begin{aligned}
\mathcal{P}_1:\quad
&\min_{\{a_{s,u}\}}
\max_{s \in \{1,\ldots,N_s\}} L_s
-
\min_{s \in \{1,\ldots,N_s\}} L_s \\[4pt]
s.t.\quad
&C1:  a_{s,u} = 0, \quad \forall u \notin \mathcal{U}_s, \\[2pt]
&C2:  a_{s,u} \in \{0,1\}, \quad
s = 1,\ldots,N_s,\;
u = 1,\ldots,N_u, \\[2pt]
&C3:  \sum_{s=1}^{N_s} a_{s,u} \le 1,
\quad \forall u \in \mathcal{U}_1, \\[2pt]
&C4:  \sum_{s=1}^{N_s} a_{s,u} \le 2,
\quad \forall u \in \mathcal{U}_2, \\[2pt]
&C5:  |\mathcal{M}_s| \le N_{\mathcal{M}_s}^{\max},
\quad |\mathcal{U}_s| \le N_{\mathcal{U}_s}^{\max}, \\[2pt]
&C6:  L_s =
\sum_{u=1}^{N_u}
\frac{a_{s,u} d_u}{\sum_{s'=1}^{N_s} a_{s',u}},
\end{aligned}
\end{equation}
where $a_{s,u}$ is a binary variable indicating whether the $u$-th user is associated with the $s$-th satellite. The constraints $C3$ and $C4$ imply the ordinary user is equipped with a single receiving antenna and the high-demand user is equipped with dual receiving antennas. In $C5$, $N_{\mathcal{M}_s}^{\max}$ and $N_{\mathcal{U}_s}^{\max}$ represent the maximum number of beam positions and users that each satellite supports. 

\subsubsection{Multi-satellite coordination BH subproblem} 
The subproblem aims to achieve an optimized trade-off among energy efficiency (EE), system throughput, and user satisfaction fairness through muli-satellite cooperative beam scheduling and onboard power allocation mechanisms. Specifically:
\begin{equation}
\label{eq4}
\begin{aligned}
\mathcal{P}_2:\;
\max_{\{ x_u^{s,k,t},\, P_s^t \}}
&\;
\omega_1 \sum_{t=1}^{T_s}
\frac{
\sum_{s=1}^{N_s} \sum_{k=1}^{K} \sum_{u=1}^{N_u}
Th_u^{s,k,t}
}{
\sum_{s=1}^{N_s} P_s^t \cdot E_{\mathrm{norm}}
}
\\
&\;+\;
\omega_2 \sum_{t=1}^{T_s}
\frac{
\sum_{s=1}^{N_s} \sum_{k=1}^{K} \sum_{u=1}^{N_u}
Th_u^{s,k,t}
}{
Th_{\mathrm{norm}}
}
\\
&\;-\;
\omega_3 \sum_{t=1}^{T_s}
\mathrm{std}\!\left(
\left\{
\frac{\bar{R}_u^{t}}{\hat{R}_u^{t}}
\right\}_{\substack{u=1,\ldots,N_u}}
\right)
\\[4pt]
\text{s.t.}\quad
&C1:\;
\sum_{u=1}^{N_u} \sum_{k=1}^{K} x_u^{s,k,t} \le K,
\quad
\forall s,\; t,
\\
&C2:\;
\eta P_{\max} \le P_s^t \le P_{\max},
\quad
\forall s,\; t,
\\
&C3:\;
\omega_1 + \omega_2 + \omega_3 = 1,
\end{aligned}
\end{equation}
where $\bar R_u^t$ is the cumulative data delivered to user $u$ by all its serving satellites up to slot $t$, and $\hat{R}_u^t$ is its cumulative traffic demand over the same period. The ratio $\bar R_u^t/\hat R_u^t$ measures service satisfaction. Its standard deviation across all users is used as the fairness metric, with lower values indicating more balanced service satisfaction. The optimization variable $x_u^{s,k,t}$ is responsible for the multi-satellite cooperative beam scheduling decision, while $P_s^t$ determines the power allocation. The constraint $C1$ indicates that each satellite serves at most $K$ beam positions simultaneously in any time slot. In $C2$, to achieve a balance between energy efficiency and system performance, the total power on board of each satellite is constrained to be no less than a prescribed lower limit $\eta P_{max}$ and no greater than its maximum available power $P_{max}$. The coupled binary beam-scheduling decisions, continuous power
control, and interference-dependent rates make $\mathcal{P}_2$ a non-convex
mixed-integer nonlinear program.

\section{User-Satellite Association and Topology-Aware Cooperative Beam Scheduling}
In subproblem $\mathcal{P}_1$, the decision variable $a_{s,u}$ is a binary variable indicating whether the $u$-th user is associated with the $s$-th satellite. Subject to the complex constraints $C2$–$C5$, the subproblem is computationally challenging due to its binary association
variables and coupled feasibility constraints. Thus, a low-complexity iterative solution is proposed to incrementally update the integer decision variables, rendering it suitable for practical satellite–user association scenarios. The detailed procedure is summarized in Algorithm~\ref{alg1}. Let $\Delta=\max_m|\mathcal{S}_m|$ denote the maximum satellite
overlap per beam position. Since $K_u\leq2$, Algorithm~1 has
complexity
$\mathcal{O}(N_u\log N_u+N_uN_s\log N_s+N_uN_s\Delta)$,
which becomes $\mathcal{O}(N_uN_s^2)$ in the worst case and
reduces to $\mathcal{O}(N_uN_s\log N_s)$ under bounded regional
overlap. Since $K_u\leq2$, Algorithm~1 terminates after at most
$\sum_u K_uN_s\leq2N_uN_s$ candidate checks. Each accepted
association satisfies the visibility, capacity, cardinality, and
angular-separation tests. Hence, the corresponding feasibility
constraints are preserved whenever sufficient feasible candidates
exist.


\begin{algorithm}[t]
\caption{An Iterative Solution to Subproblem $\mathcal{P}_1$}
\label{alg1}
\small
\begin{algorithmic}[1]
\REQUIRE
User set $\mathcal{U}$, satellite set $\mathcal{S}$, user traffic demand $\{d_u\}$, and high-demand user set $\mathcal{I}$.
\ENSURE
Satellite-user association indicator $a_{s,u}$.

\STATE Sort users in descending order of
$d_u/[1+\mathbb{I}(\sum_{s=1}^{N_s}a_{s,u}>1)]$.
\FOR{$u=u_1$ to $u_{N_u}$}
    \STATE Sort satellites in ascending order of load $L_s$.
    \STATE$K_u \leftarrow 2$ if $u\in\mathcal{I}$; otherwise $K_u \leftarrow 1$; \quad $count\leftarrow 0$.
    
    \FOR{$r=1$ to $K_u$} 
        \FOR{$s=s_1$ to $s_{N_s}$}
            \IF{$count=K_u$} \STATE \textbf{break} \ENDIF
            \IF{$\neg\mathrm{Vis}(s,u)\ \lor\ |\mathcal{U}_s|\!\ge\! N_{U_s,\max}\ \lor\ |\mathcal{M}_s|\!\ge\! N_{M_s,\max}$}
                \STATE \textbf{continue}
            \ENDIF
            \IF{$s\notin\mathcal{S}_{m(u)}$ \AND $\exists\,s'\!\in\!\mathcal{S}_{m(u)}:\angle(s,s';m(u))<\theta_{\min}$}
                \STATE \textbf{continue}
            \ENDIF
            \STATE $a_{s,u}\!\leftarrow\!1$;\  $\mathcal{M}_s\!\leftarrow\!\mathcal{M}_s\cup\{m_u\}$;
                $\mathcal{S}_{m_u}\!\leftarrow\!\mathcal{S}_{m_u}\cup\{m_u\}$;\quad$count\!\leftarrow\!count+1$.
        \ENDFOR
    \ENDFOR
\ENDFOR
\end{algorithmic}
\end{algorithm}

The subproblem $\mathcal{P}_2$ is modeled as a partially observable Markov decision process (POMDP), where each satellite functions as an autonomous agent with partial observations. The state, observation, action, and reward are defined as follows:
\begin{itemize}
\item {\em State}:
The environment state consists of the global user service states, channel states, and location information associated with all satellites, which can be expressed as:
\begin{equation}
\label{eq6}
\begin{array}{l}
{{\cal S}^t} = \{ {Q_{s,{u_s}}}(t),{H_{s,{u_s}}}(t),{L_{s,{u_s}}}(t)|\\
s = 1,2,...,{N_s},{u_s} \in {{\cal U}_s}\} 
\end{array}.
\end{equation}
The traffic state $Q_{s,{u_s}}(t)$ includes the on-board traffic load, the average delay, the delay in the head-of-line (HoL), the arrival rate of the traffic, average throughput, and the initial demand for service. The user location information ${L_{s,{u_s}}}(t)$ comprises the latitude and longitude coordinates of the serving beam, the associated satellite index, and the remaining service time. 
\item {\em Observation}:
In the considered scenario, each satellite observes only the local information of its served users. Thus, the observation of the $s$-th satellite is given by
\begin{equation}
\label{eq7}
{\cal O}^t_s = \{ {Q_{s,{u_s}}}(t),{H_{s,{u_s}}}(t),{L_{s,{u_s}}}(t)|{u_s} \in {{\cal U}_s}\}.
\end{equation}

\item {\em Action}:
The action of each agent corresponds to the beam scheduling and on-board power control decisions, i.e., $a^t_s = \{ X^t_s,P^t_s\} ,X^t_s = {\{ x_u^{s,k,t}\} _{u \in {{\cal U}_s}}}$.

\item {\em Reward}:
The instantaneous reward $r^t$ is defined as
\begin{equation}
\label{eq8}
\begin{aligned}
r_t =\;& \omega_1 
\frac{\sum\limits_{s=1}^{N_s}\sum\limits_{k=1}^{K}\sum\limits_{u=1}^{N_u} Th_{u}^{s,k,t}}
{\sum\limits_{s=1}^{N_s} P^t_s \cdot \mathbb{E}\{E_{\mathrm{norm}}\}}
+ \omega_2 
\frac{\sum\limits_{s=1}^{N_s}\sum\limits_{k=1}^{K}\sum\limits_{u=1}^{N_u} Th_{u}^{s,k,t}}
{Th_{\mathrm{norm}}} \\
&- \omega_3 \,
\mathrm{std}\!\left\{
\frac{\bar{R}_u^{t}}{\hat{R}_u^{t}}
\right\}_{\substack{u=1,2,\dots,N_u}} .
\end{aligned}
\end{equation}
\end{itemize}

Considering the limited capability for real-time inter-satellite communication and the partially observable environment, the MAPPO algorithm under the CTDE framework is adopted. During training, MAPPO employs a centralized critic for value estimation and updates the policy using the clipped surrogate objective of PPO, which has been widely shown to ensure stable learning in cooperative multi-agent tasks. The policy network of the $s$-th agent is defined as
\begin{equation}
\label{eq9}
a_s^t
\sim
\pi_{\theta}\left(\cdot \mid h_s^t\right), \\
h_s^t
=
\operatorname{GRU}
\left(h_s^{t-1},\widetilde{o}_s^t\right), \\
\widetilde{o}_s^t
=
\left[o_s^t \,\Vert\, F_s^{(e)}\right],
\end{equation}
where \(\widetilde{o}_s^t\) is the topology-augmented observation, \(h_s^t\) is the recurrent hidden state, and \(F_s^{(e)}\) is the episode-level topology representation. The two message-passing phases capture complementary dependencies. Phase I aggregates the satellite attributes with the traffic and link features of its associated users to obtain the local service embedding \(\mathbf{z}_s^{(e)}\). Since these embeddings are generated independently, Phase I alone cannot capture inter-satellite coupling. Phase II propagates them through the association-induced matrix \(\mathbf{A}_{ss}^{(e)}=(\mathbf{A}_{su}^{(e)})^{T}\mathbf{A}_{su}^{(e)}\), incorporating cross-satellite structural relations. Therefore, omitting Phase II leaves only isolated local summaries, whereas omitting Phase I removes user- and link-conditioned information. Both phases are necessary to compactly encode local service context and cross-satellite coupling. \(F_s^{(e)}\) remains fixed within each episode.
\begin{itemize}

\item \textit{Satellite--User Graph Construction:}
At the beginning of episode $e$, the user--satellite association
induces a heterogeneous graph
$G^{(e)}=(V_u^{(e)}\cup V_s^{(e)},E^{(e)})$. The user-node feature
$\mathbf{x}_u^{(e)}$ characterizes the traffic arrival rate, while the
satellite-node feature $\mathbf{x}_s^{(e)}$ contains its latitude,
longitude, and orbital altitude.
An edge $(u,s)\in E^{(e)}$ exists if satellite $s$ is visible to user $u$, and its
feature is the channel gain evaluated at episode initialization,
i.e., $\mathbf{e}_{u,s}^{(e)}=g_{u,s}^{(e)}
=|h_{u,s}^{(e,0)}|^2$.

\item \textit{Two-phase Message Passing:}
In the first phase, the user and link features are aggregated at each
serving satellite:
\begin{equation}
\mathbf{z}_{s}^{(e)}
=
\mathrm{LN}\{
\phi_{s}(\mathbf{x}_{s}^{(e)})
+
\phi_{o}[
\textstyle\sum_{u\in\mathcal{U}_{s}^{(e)}}
\alpha_{u,s}^{(e)}
\phi_{us}(\mathbf{x}_{u}^{(e)},\mathbf{e}_{u,s}^{(e)})
]\},
\label{eq:user_sat_aggregation}
\end{equation}
where $\mathrm{LN}(\cdot)$ denotes layer normalization;
$\phi_s$, $\phi_{us}$, and $\phi_o$ are the satellite projection,
user--edge message, and update mappings, respectively; and
$\alpha_{u,s}^{(e)}$ is a sigmoid gate. 

In the second phase, the weighted adjacency matrix
$A_{su}^{(e)}$, with
$[A_{su}^{(e)}]_{u,s}=a_{s,u}^{(e)}g_{u,s}^{(e)}$, induces the
satellite structural matrix $A_{ss}^{(e)}
=
\left(A_{su}^{(e)}\right)^{\mathsf{T}}
A_{su}^{(e)}$. The topology-aware representation is then obtained as
\begin{equation}
\mathbf F_{s}^{(e)}
=
\mathrm{LN}\{
\mathbf z_{s}^{(e)}
+
\textstyle\sum_{s'=1}^{N_s}
\widetilde a_{s,s'}^{(e)}
\phi_{ss}(\mathbf z_{s'}^{(e)})
\},
\label{eq:sat_sat_aggregation}
\end{equation}
where $\widetilde a_{s,s'}^{(e)}$ is the $(s,s')$th entry of
$\widetilde{\mathbf A}_{ss}^{(e)}$, and $\phi_{ss}$ is the
satellite-message mapping. 
\end{itemize}
\begin{figure}[!t]
\centering
\includegraphics[width=0.7\columnwidth]{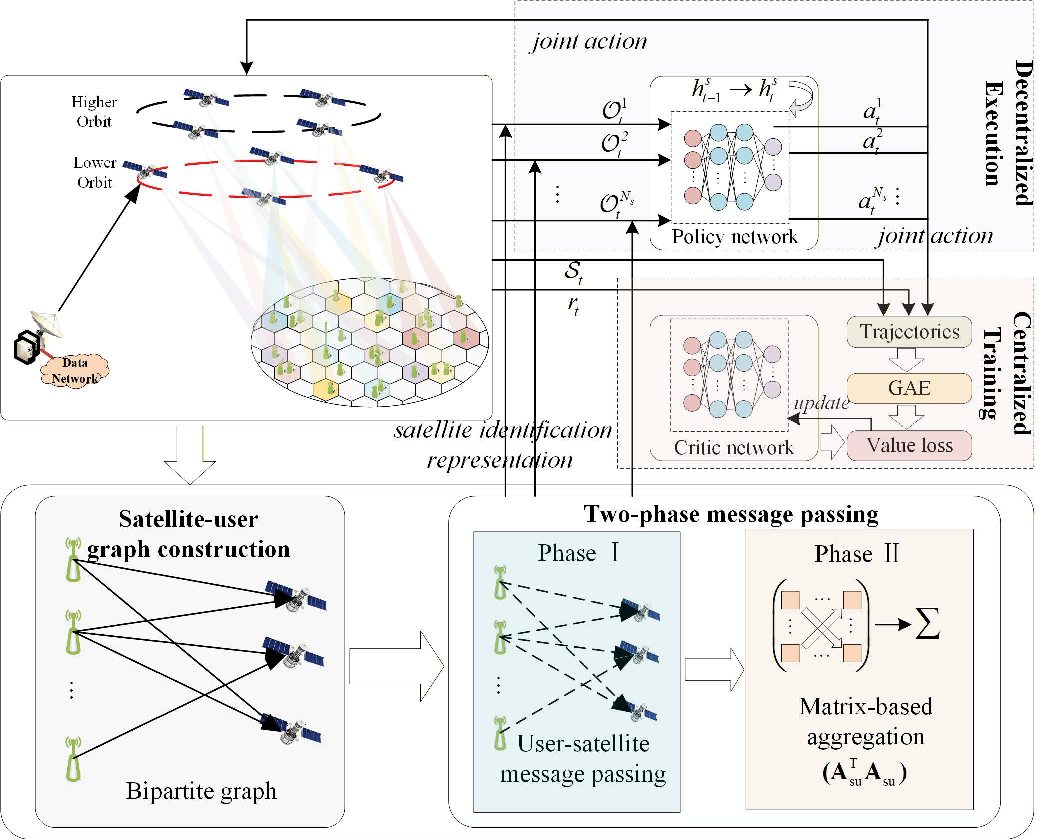}
\vspace{-6mm}
\caption{Algorithm architecture.}
\label{fig_2}
\end{figure}
\begin{algorithm}[t]
\caption{TA-MAPPO Algorithm}
\label{alg2}
\small
\begin{algorithmic}[1]
\STATE Initialize Policy and Critic network parameters $\theta$, $\phi$.
\STATE Initialize replay buffer $\mathcal{D}$.
\FOR{each episode $e=1,2,\dots,N_{\text{eps}}$}
    \STATE Reset environment and obtain initial state $\mathcal{S}^0$.
    \STATE Construct graph ${{\mathcal{G}}^{(e)}} $.
    \STATE  Generate ${F_s^{(e)}}$ via two-phase message passing.
    \FOR{each time slot $t=1,2,\dots,T$}
        \FOR{each agent $s$}
            \STATE Observe local state $\mathcal{O}^t_s$.
            \STATE Generate action $a^t_s$ using the shared policy.
        \ENDFOR
        \STATE Execute joint action $\mathbf{a}^t=\{a^t_1,\dots,a^t_{N_s}\}$.
        \STATE Receive reward $r^t$ and next state $\mathcal{S}^{t+1}$.
        \STATE Store transition $(\mathcal{S}^t,\mathbf{a}^t,r^t,\mathcal{S}^{t+1})$ in trajectories.
    \ENDFOR
    \STATE Estimate state value $V(\mathcal{S}^t)$ using centralized critic.
    \STATE Compute advantage $A^t$ using GAE.
    \STATE Update parameters $\phi$ by minimizing value loss.
    \STATE Update parameters $\theta$ using clipped surrogate objective.
    \STATE Update old policy parameters $\theta^{-} \leftarrow \theta$.
\ENDFOR

\end{algorithmic}
\end{algorithm}

Algorithm~2 summarizes the training procedure of the proposed
topology-aware MAPPO (TA-MAPPO) algorithm. 
Let $E=|\mathcal{E}^{(e)}|$. With embedding width $H$, actor
width $H_{\pi}$, and observation/action dimensions $D_o/D_a$,
the graph encoder costs
$\mathcal{O}((E+N_s)H^2+N_uN_s^2+N_s^2H)$ once per episode,
while the shared recurrent actor costs
$\mathcal{O}\!\left((D_o+H)H_{\pi}+H_{\pi}^{2}
+H_{\pi}D_a\right)$ per slot. Since the actors run in parallel
with shared parameters, the per-satellite inference cost is
independent of $N_s$. Sigmoid gating, row normalization,
residual--LN updates, and the shallow architecture help stabilize
training and mitigate over-smoothing. The topology encoder only augments the policy and value inputs without modifying MAPPO's objective or update rule. Thus, the
proposed method inherits the convergence conditions and limitations
of standard MAPPO.
\section{Numerical Results}

\begin{table}[t]
\centering
\caption{Simulation Parameters}
\small   
\begin{tabular}{p{3.3cm} c}
\hline
\textbf{Parameter} & \textbf{Value} \\
\hline
Orbit 1 (higher) & 600 km / 50$^\circ$ \\
Orbit 2 (lower) & 508 km / 55$^\circ$ \\
Region & [116.25E, 118.25E], [29N, 32N] \\
Carrier frequency & Ka-band (20 GHz) \\
System Bandwidth & 200 M \\
Maximum power $P_{max}$ & 250 W \\
Number of subchannels & 20 \\
$N_{U_s,\max}$ & 80 \\
$N_{M_s,\max}$ & 60 \\
Number of beams $K$ & 16 \\
Rx antenna gain (max) & 20 dBi \\
Tx antenna gain (max) & 38.5 dBi \\
Power coefficient $\eta$ & 0.8 \\
\hline
\end{tabular}
\label{tab1}
\vspace{-5mm}
\end{table}

We simulate a Ka-band multi-satellite BH system with a GW-style
two-layer LEO constellation \cite{9} comprising eight satellites.
Each satellite supports up to 80 users, 60 beam positions, and 16
simultaneous beams, with 250~W onboard power and 200~MHz bandwidth
\cite{10}. Traffic arrivals follow a Poisson process whose rate
reflects population density and economic activity. The main
simulation parameters are listed in Table~\ref{tab1}.
The one-block GNN uses $H=64$, with user/satellite/edge projections
of $1$--$16$--$32$, $3$--$16$--$64$, and $1$--$16$--$32$,
respectively. Its message/update MLPs are $64$--$64$--$64$ with
ReLU, and the gate is a $64$--$1$ sigmoid layer. The shared actor
uses a $512$--$256$ feature extractor followed by a 256-dimensional
GRU.
\begin{figure}[!h]
\centering
\includegraphics[width=0.55\columnwidth]{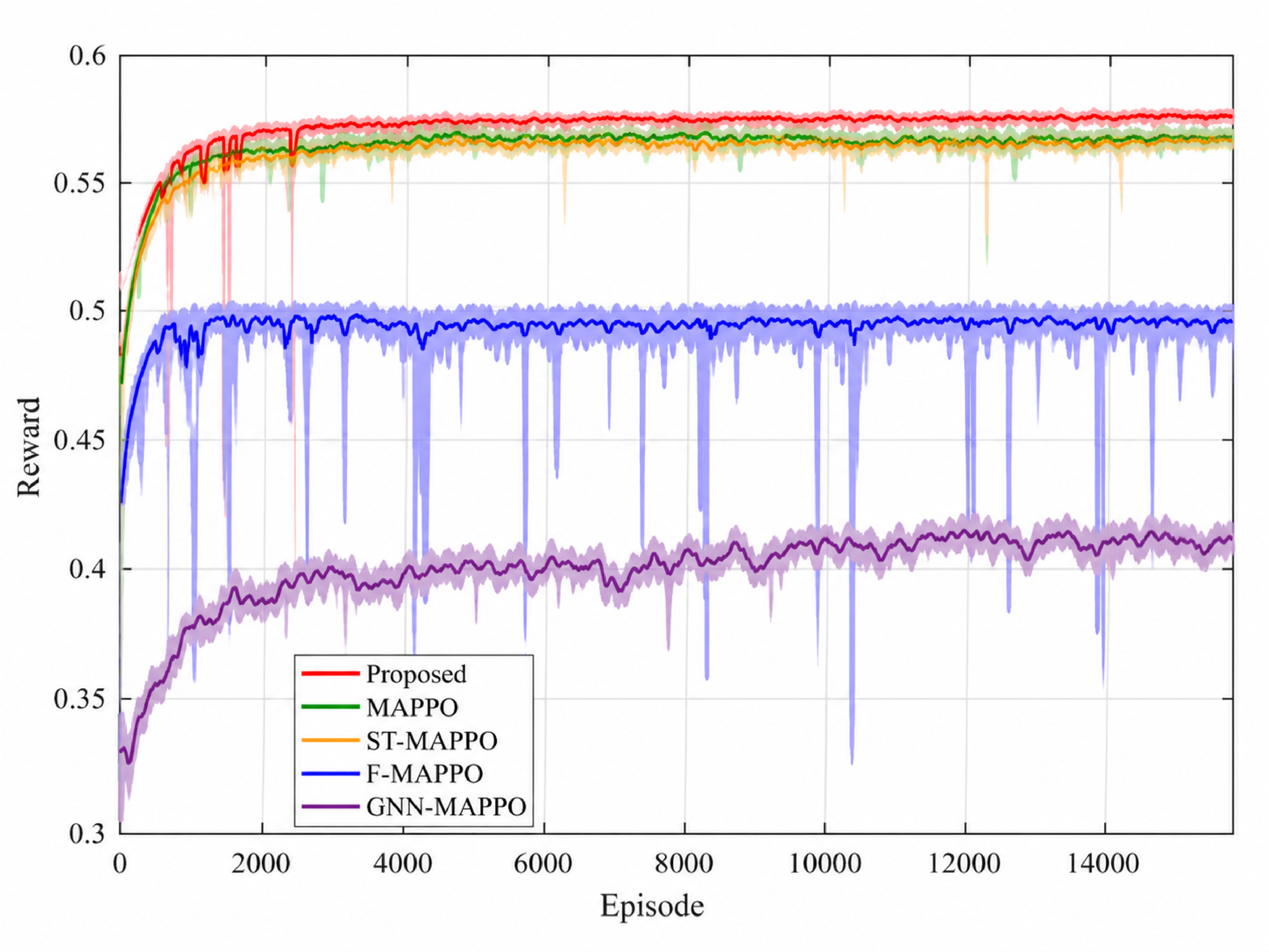}
\vspace{-4mm}
\caption{Convergence performance comparison with other DRL.}
\label{fig_3}
\vspace{-5mm}
\end{figure}

Fig. \ref{fig_3} shows the evolution of the reward combining EE, throughput, and fairness. All CTDE-based methods gradually converge. F-MAPPO, which performs beam scheduling with full-power allocation, converges to a much lower reward, highlighting the benefit of adaptive power control. MAPPO performs joint beam scheduling and power control without topology enhancement, while ST-MAPPO further incorporates a simplified topology considering only satellite positions. However, ST-MAPPO achieves nearly the same reward as MAPPO, indicating that such coarse topology provides no effective improvement. In contrast, the proposed scheme consistently achieves the highest reward, demonstrating that the GNN-based topology enhancement can exploit richer structural information for more effective joint optimization.

\begin{figure}[!t]
\centering
\includegraphics[width=0.65\columnwidth]{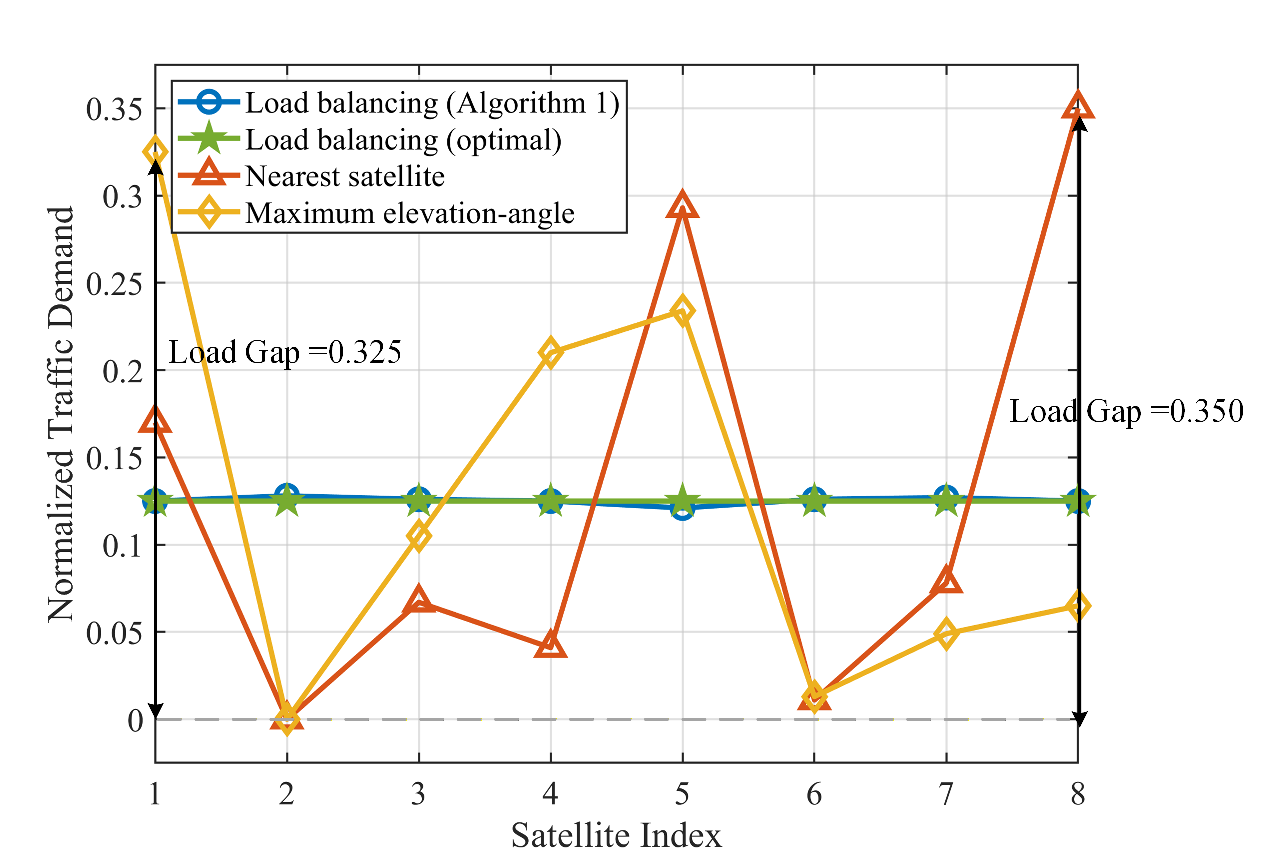}
\vspace{-4mm}
\caption{Results of load balancing.}
\label{fig_4}
\vspace{-3mm}
\end{figure}

We further compare the proposed load-balancing Algorithm~\ref{alg1}
with the nearest-satellite, maximum-elevation-angle, and SCIP-based
optimal association strategies. As shown in Fig.~\ref{fig_4}, the
two conventional schemes yield load gaps of 0.350 and 0.325,
whereas Algorithm~\ref{alg1} achieves an almost uniform load
distribution closely matching the SCIP optimum. This confirms a
small optimality loss with substantially lower computational cost. Table \ref{tab2} summarizes the performance comparison in terms of load gap, EE, throughput, and fairness for the three user-satellite association strategies. The results indicate that the proposed load-balanced association significantly outperforms the other two. Specifically, energy efficiency increases by 57.14\% and 48.42\% and throughput improves by 79.27\% and 69.52\%. The standard deviation of service satisfaction ratios decreases from 0.41 for both conventional association strategies to 0.36, indicating more balanced service satisfaction.

\begin{table}[t]
\centering
\footnotesize
\setlength{\tabcolsep}{4pt}
\renewcommand{\arraystretch}{1.05}
\caption{Performances of User–Satellite Association Methods}
\begin{tabular}{|c|c|c|c|c|}
\hline
\makecell{\textbf{Association}\\\textbf{strategies}} &
\makecell{\textbf{Load gap}\\(Gbps)} &
\makecell{\textbf{EE}\\(Mbps/W)} &
\makecell{\textbf{Throughput}\\(Gbps)} &
\textbf{Fairness(std.)} \\
\hline
Load-balancing          & 0.21  & 13.64 & 21.84 & 0.36 \\
\hline
Nearest satellite       & 11.49 & 8.68  & 12.18 & 0.41 \\
\hline
\makecell{Maximum\\elevation-angle} & 10.76 & 9.19 & 12.88 & 0.41 \\
\hline
\end{tabular}
\label{tab2}
\vspace{-5mm}
\end{table}

\begin{figure}[!t]
\centering
\includegraphics[width=\columnwidth]{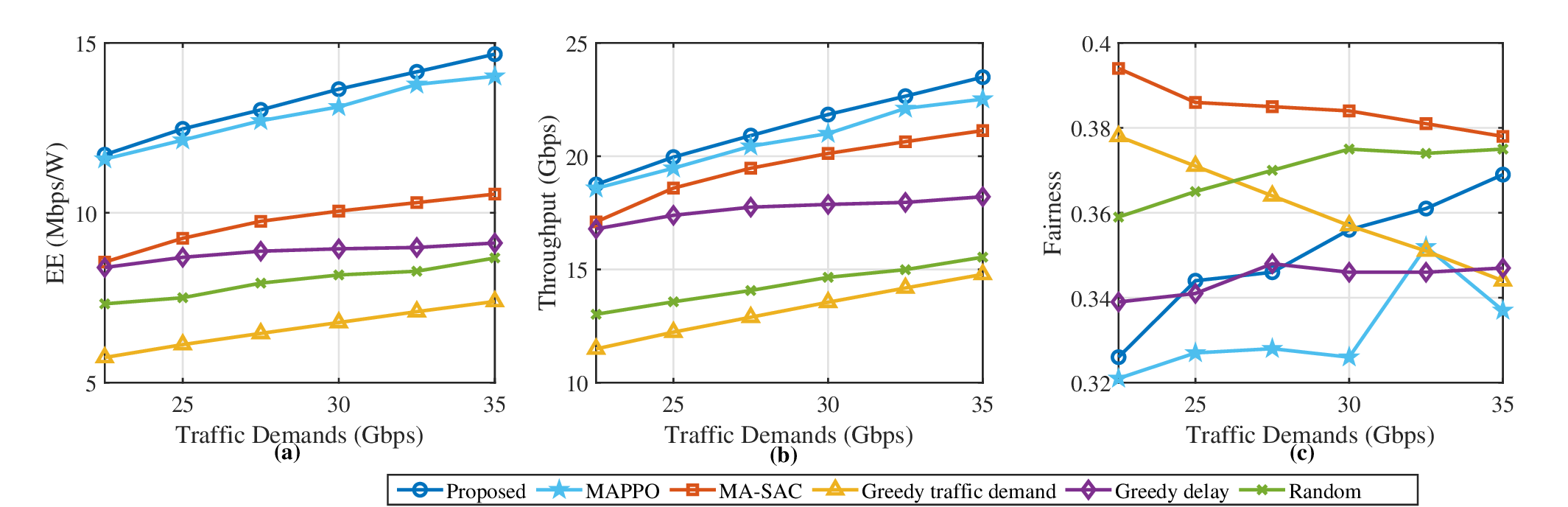}
\vspace{-5mm}
\caption{Performances of different algorithms vs. Traffic demands.}
\label{fig_5}
\vspace{-5mm}
\end{figure}

The baseline methods 
are summarized as follows:

\begin{itemize}
\item \textbf{MA-SAC}: A CTDE-based approach that employs the Soft Actor–Critic (SAC) algorithm for beam scheduling, with satellites operating at maximum transmission power.
\item \textbf{Greedy traffic demand}: Each satellite operates at full power and allocates its beams to the $K$ beam spots with the highest traffic demand.
\item \textbf{Greedy delay}: Full-power operation, prioritizing the 
$K$ beam positions with the largest waiting delay.
\item \textbf{Random}: Each satellite randomly selects 
$K$ beam positions and randomly allocates power.
\end{itemize}

Fig.~\ref{fig_5} compares the methods under different traffic demands, where MAPPO omits the topology-aware representation but retains the same cooperative control framework. The proposed scheme consistently achieves higher EE and throughput than MAPPO as traffic load increases, demonstrating the effectiveness of topology-aware enhancement. It also outperforms MA-SAC and the heuristic schemes overall while maintaining competitive fairness. Greedy delay maintains relatively stable fairness by prioritizing waiting delay.

\section{Conclusions}
This letter developed a topology-aware cooperative beam-hopping (BH) framework for multi-satellite systems. The load-balanced serving relationships induce an episode-specific satellite–user graph, from which two-phase message passing generates topology-aware structural identities for parameter-sharing agents. Combined with slot-level local observations, these representations support decentralized joint beam scheduling and onboard power control under the CTDE framework. Simulation results demonstrated improvements in energy efficiency,
throughput, and fairness. Future work will investigate larger service regions, cross-region coordination, imperfect CSI, and hardware-level latency
validation.


\begin{thebibliography}{10}

\bibitem{1}
X.~Ma, H.~Zhang, W.~Song, Y.~Lu, Y.~Wu, and V.~C.~M. Leung,
``Resource Optimization for LEO Constellation Networks: A Multi-Satellite Cooperative Coverage Design,''
\textit{IEEE Trans. Commun.}, 
vol.~73, no.~11, pp.~10201--10216, 2025.

\bibitem{2}
S.~Yuan, Y.~Sun, M.~Peng, and R.~Yuan,
``Joint beam direction control and radio resource allocation in dynamic multi-beam LEO satellite networks,''
\textit{IEEE Trans. Veh. Technol.},
vol.~73, no.~6, pp.~8222--8237, 2024.

\bibitem{3}
Z.~Lin, Z.~Ni, L.~Kuang, C.~Jiang, and Z.~Huang,
``Multi-satellite beam hopping based on load balancing and interference avoidance for NGSO satellite communication systems,''
\textit{IEEE Trans. Commun.},
vol.~71, no.~1, pp.~282--295, 2022.

\bibitem{4}
Z.~Lin, Z.~Ni, L.~Kuang, C.~Jiang, and Z.~Huang,
``Satellite-Terrestrial Coordinated Multi-Satellite Beam Hopping Scheduling Based on Multi-Agent Deep Reinforcement Learning,''
\textit{IEEE Trans. Wireless Commun.}, 
vol.~23, no.~8, pp.~10091--10103, 2024.

\bibitem{5}
R.~Zhao, J.~Cai, J.~Luo, Y.~Ran, J.~Gao, and Y.~Xu,
``Joint Beam Hopping and Resource Allocation for Load Balancing and Interference Avoidance in Multi-LEO Satellite Networks,''
in \textit{Proc. IEEE ICC}, 2025, pp.~958--963.

\bibitem{6}
X.~Xie, K.~Fan, W.~Deng, N.~Pappas, and Q.~Zhang,
``Multi-Satellite Beam Hopping and Power Allocation Using Deep Reinforcement Learning,''
\textit{arXiv preprint arXiv:2501.02309}, 2025.

\bibitem{7}
F.~Hu, X.~Zhuo, L.~Tang, Y.~Zhao, Z.~Bu, and F.~Qu,
``Multi-Objective Dynamic Beam Hopping Scheme based on MADDPG in LEO Satellite Communication System,''
in \textit{Proc. IEEE INFOCOM Workshops}, 2025, pp.~1--6.

\bibitem{8}
C.~Ren, F.~Tian, Y.~Zhao, W.~Wang, Z.~Yang, Y.~Gao, and H.~Liu,
``A Multi-Agent Reinforcement Learning Based Cooperative Beam Hopping Scheme for LEO Mega Constellation Networks,''
in \textit{Proc. IEEE WCNC}, 2025, pp.~1--6.

\bibitem{9}
Y.~Xu, Z.~Wang, X.~Hu, and Y.~He,
``Self-Organizing Maintenance Method of Mega Walker Constellations With Multiple Coverage,''
\textit{IEEE Trans. Aerosp. Electron. Syst.},
pp.~1--17, 2025.

\bibitem{10}
J.~Zhang, D.~Qin, C.~Kong, F.~Zhao, R.~Li, J.~Wang, and Y.~Wang,
``System-level evaluation of beam hopping in NR-based LEO satellite communication system,''
in \textit{Proc. IEEE WCNC}, 2023, pp.~1--6.

\end{thebibliography}
\end{document}